\documentclass[twocolumn]{aastex61}

\shorttitle{Influence of sub-stellar land on exoplanet climate}
\shortauthors{Lewis et al.}

\received{October 17, 2017}
\revised{January 29, 2018}
\accepted{February 1, 2018}
\submitjournal{ApJ}

\begin{document}

\title{The influence of a sub-stellar continent on the climate of a tidally-locked exoplanet}

\correspondingauthor{Neil T. Lewis}
\email{nl290@exeter.ac.uk}

\author{Neil T. Lewis}
\affiliation{Mathematics, College of Engineering, Mathematics and Physical Sciences, University of Exeter, Exeter, EX4 4QF, UK}
\affiliation{Physics and Astronomy, College of Engineering, Mathematics and Physical Sciences, University of Exeter, Exeter, EX4 4QL, UK}

\author{F. Hugo Lambert}
\affiliation{Mathematics, College of Engineering, Mathematics and Physical Sciences, University of Exeter, Exeter, EX4 4QF, UK}

\author{Ian A. Boutle}
\affiliation{Met Office, FitzRoy Road, Exeter, EX1 3PB, UK}
\affiliation{Physics and Astronomy, College of Engineering, Mathematics and Physical Sciences, University of Exeter, Exeter, EX4 4QL, UK}

\author{Nathan J. Mayne}
\affiliation{Physics and Astronomy, College of Engineering, Mathematics and Physical Sciences, University of Exeter, Exeter, EX4 4QL, UK}

\author{James Manners}
\affiliation{Met Office, FitzRoy Road, Exeter, EX1 3PB, UK}
\affiliation{Physics and Astronomy, College of Engineering, Mathematics and Physical Sciences, University of Exeter, Exeter, EX4 4QL, UK}

\author{David M. Acreman}
\affiliation{Physics and Astronomy, College of Engineering, Mathematics and Physical Sciences, University of Exeter, Exeter, EX4 4QL, UK}
\affiliation{Computer Science, College of Engineering, Mathematics and Physical Sciences, University of Exeter, Exeter, EX4 4QF, UK}

\begin{abstract}

Previous studies have demonstrated that continental carbon-silicate weathering is important to the continued habitability of a terrestrial planet. Despite this, few studies have considered the influence of land on the climate of a tidally-locked planet. In this work we use the Met Office Unified Model, coupled to a land surface model, to investigate the climate effects of a continent located at the sub-stellar point. We choose to use the orbital and planetary parameters of Proxima Centauri B as a template, to allow comparison with the work of others. A region of the surface where $T_{\text{s}} > 273.15\,\text{K}$ is always retained, and previous conclusions on the habitability of Proxima Centauri B remain intact.  We find that sub-stellar land causes global cooling, and increases day-night temperature contrasts by limiting heat redistribution. Furthermore, we find that sub-stellar land is able to introduce a regime change in the atmospheric circulation. Specifically, when a continent offset to the east of the sub-stellar point is introduced, we observe the formation of two mid-latitude counterrotating jets, and a substantially weakened equatorial superrotating jet.

\end{abstract}


\section{Introduction} \label{sec:intro}

Beginning with the works of \citet{1997Icar..129..450J} and \citet{2002A&A...385..166S}, 3D atmosphere general circulation models have been used to study the character of planets beyond our solar system. The output from 3D models has uncovered new dynamical regimes \citep[see][for a review]{2010exop.book..471S}, and has allowed characterization of the `habitability' of recently discovered \emph{terrestrial} exoplanets orbiting M-Dwarf stars such as Proxima Centauri B (\citealt{2016A&A...596A.112T,2017A&A...601A.120B,2017arXiv170902051D}, discovery: \citealt{2016Natur.536..437A}) and the Trappist-1 planets (\citealt{2017ApJ...839L...1W,2017arXiv170706927T}, discovery: \citealt{2017Natur.542..456G})

M-dwarf stars are believed to make up approximately 75\% of stars on the main sequence. Using a conservative definition for the habitable zone, \citet{2015ApJ...807...45D} estimate there are approximately $0.16$ Earth size planets, and $0.12$ super-Earth size planets per M-dwarf habitable zone. As a result, habitable zone planets orbiting M dwarf stars are expected to be numerous. M-dwarf stars are much smaller and cooler, and subsequently dimmer, than the Sun, making the potential to detect habitable planets orbiting these stars much improved over Sun-like stars. The reduced stellar radii of M-dwarfs compared to that of G-dwarfs leads to a stronger per-transit signal. Moreover, in order to achieve potentially habitable temperatures planets must orbit M-dwarfs much more closely than they would around a G-dwarf, leading to much shorter orbital periods and increased feasibility of repeat observations further increasing the attraction of these targets. Due to strong tidal forces, planets orbiting at such short distances may fall into so-called `tidally-locked' synchronous rotation, where the planet's orbital period and rotation period are the same. This means that the planet has a permanent `day-side' and a permanent `night-side'. 3D climate modeling studies of tidally-locked planets have shown that the day-night forcing difference experienced by these planets gives rise to atmospheric circulation and equilibrium climate different to that found on Earth  \citep[e.g.][]{2010JAMES...2...13M,2013A&A...554A..69L,2015ApJ...804...60K}.

Traditionally, the habitability of a planet has been assessed by whether it occupies the `habitable zone' of its parent star. The habitable zone is defined as the area around a star within which a planet can orbit and host liquid water at its surface \citep{1993Icar..101..108K}. Its inner boundary is determined by a climate transition to either a `moist greenhouse' or `runaway greenhouse' state. In the former scenario, water in the stratosphere is lost by photolysis and hydrogen is lost to space. In the latter, the outgoing thermal radiation reaches an upper limit beyond which surface temperatures can rise unchecked, until a planet's oceans are evaporated into the atmosphere \citep{2016ApJ...819...84K}. The outer boundary of the habitable zone is traditionally defined as the point after which surface temperatures are cool enough to allow $\text{CO}_{2}$ to condense onto the surface, removing the $\text{CO}_{2}$ greenhouse effect, thus causing global cooling and consequently global glaciation. The outer boundary can be extended if greenhouse gasses such as hydrogen are present in sufficient quantity to forgo the requirement for $\text{CO}_{2}$ greenhouse warming \citep{ 2011ApJ...734L..13P}. 

Recently, some studies have advanced the assessment of habitability by systematically assessing the \emph{different} equilibrium states an individual planet could occupy. For example, \citet{2011ApJ...726L...8P} and \citet{2016A&A...596A.112T} investigate the effects of different possible atmospheric compositions on the climates of Gliese 518g \citep[discovery:][]{2010ApJ...723..954V} and Proxima Centauri b (hereafter, ProC b), respectively. These studies are informed by discussion regarding the initial volatile inventory of the planet and water-loss as a result of interaction with the planet's host star \citep[for a discussion on the evolution of ProC b's volatile inventory, see][]{2016A&A...596A.111R}. \citet{2017A&A...601A.120B} discuss the effects of ProC b occupying a 3:2 resonant orbit with its host star, as opposed to a tidally-locked orbit on planetary climate and habitability.

To remain habitable on the order of gigayears it has been suggested that a planet requires a carbon-silicate weathering cycle, whereby it can sequester carbon in order to balance volcanic outgassing of $\text{CO}_{2}$ and `adapt' to rising temperatures resulting from increased stellar irradiance as a star progresses through its main sequence life-time \citep{1993Icar..101..108K}. Continued habitability on the order of gigayears is important, as life may only emerge on a planet a few hundred million years subsequent to the planet's formation. The earliest confirmed evidence for life on Earth is found to be from the Archaean, 3.5 Gyr ago, and it is thought therefore that life began at the end of the Hadean or in the early Archaean, roughly 500 Myr - 1 Gyr after the formation of the Earth \citep{2001Natur.409.1083N}.

Silicate weathering can occur either on continents or at the sea floor.  Continental weathering is facilitated by precipitation, which contains dissolved $\text{CO}_{2}$ that reacts with silicate rocks to create aqueous minerals. These minerals are then transported by surface water run-off to the ocean, where carbonates are created and subsequently buried. An increase in temperature would cause an increase in precipitation, thus increasing weathering and the $\text{CO}_{2}$ draw down rate. This, in turn, would reduce the $\text{CO}_{2}$ greenhouse effect, reducing the temperature. In this way, carbon-silicate weathering provides a stabilizing feedback that allows a planet to respond to increased insolation. For an extended description of the carbon-silicate weathering cycle the reader is invited to consult \citet{2012ApJ...756..178A}. 

Sea floor weathering is thought to be weaker than continental weathering, and far less temperature dependent \citep{2012ApJ...756..178A}. Assuming sea floor weathering is temperature independent, \citet{2012ApJ...756..178A} find that ocean planets without a continental surface (hereafter, aquaplanets) will be unable to adapt to increased stellar irradiance, and therefore the habitable zone for such planets will be `dramatically narrower' than previously thought. \citet{2015MNRAS.452.3752K} note that aquaplanets may host deep oceans, with depth in excess of $100\,\text{km}$. The pressure associated with such depth will cause the formation of `high-pressure water ice' at the ocean floor, which could prevent carbon exchange with the continental crust below. This would compromise the ability of the carbon-silicate cycle to sequester carbon. With this in mind, and in light of the findings presented by \citet{2012ApJ...756..178A}, they suggest that the rate of atmospheric $\text{CO}_{2}$ draw down will be controlled by the ability of the ocean to dissolve atmospheric $\text{CO}_{2}$, as opposed to by the carbon-silicate cycle. In this scenario, where oceanic $\text{CO}_{2}$ dissolution is the primary mechanism for carbon exchange with the atmosphere, they find that the carbon cycle may actually provide a positive feedback, and thus a destabilizing effect on climate. This would serve to reduce the width of the habitable zone. It follows from the results of \citet{2012ApJ...756..178A} and \citet{2015MNRAS.452.3752K} that an exposed continental surface may be required for a planet to maintain an effective carbon-silicate cycle. Indeed, \citet{2012ApJ...756..178A} show that if a land mass is present then continental weathering can occur, with little dependence on land fraction, allowing us to retain previous habitable zone theory and limits. 

It seems sensible to consider the climate dynamics of tidally-locked planets where a continent is present, yet to date few studies have done so, with the majority choosing instead to focus on aquaplanets or entirely land planets. \citet{2003AsBio...3..415J} and \citet{2017arXiv170902051D} present results for simulations where continents are introduced, but as part of broader studies into planetary climate, and so few results presented are continent specific. The aim of this study is to investigate the climate response to the introduction of a continent on a tidally-locked planet. We choose to focus on Proxima Centauri b, and assume it has an $\text{N}_{2}$ dominated atmosphere with a surface pressure of $p_{s} = 10^{5}\,\text{Pa}$  and trace $\text{CO}_{2}$ to allow easy comparison with \citet{2017A&A...601A.120B} and \citet{2016A&A...596A.112T}. 

As discussed in \citet{2016A&A...596A.112T}, it has been suggested for tidally-locked planets that any large-scale gravitational anomaly, and by extension, topographical anomaly, is likely to be aligned with the star-planet axis \citep{Wieczorek2007}. Therefore, one might expect a topographical basin to be located at either the anti-stellar point, or at the sub-stellar point, as is the case on the Moon \citep{1994Sci...266.1839Z}. This could favour the existence of any above- sea-level land at the sub- or anti-stellar point. Continental weathering requires precipitation, which falls largely near the sub-stellar point, and is only likely to be effective when land is not covered in ice, which requires $T_{\text{s}}>273.15\,\text{K}$  as is the case on the day-side of our planet \citep{2017A&A...601A.120B}. Indeed, \citet{2012AsBio..12..562E} demonstrate that carbon-silicate weathering is greatly enhanced when land is located on the day-side of a tidally-locked planet, as opposed to the night-side. Furthermore, it is at the sub-stellar point that we expect land to cause the largest change in climate from the aquaplanet scenario. With these points in mind, for this study we consider land located at the sub-stellar point.

A sub-stellar continent will throttle the supply of moisture to the atmosphere, reducing the atmospheric temperature and water vapor, and the occurrence of cloud and precipitation.  Notwithstanding reduction in cloud, these effects are expected to reduce surface temperatures globally through a reduction in the water vapor greenhouse effect, and to reduce the efficiency of moist atmospheric heat transport. On a rapidly, non-synchronously rotating planet such as Earth, a drier atmosphere, and thus a reduced capacity for moist atmospheric heat transport, serves to reduce meridional heat transport, cooling polar regions \citep[as presented in][their Figure 4]{2013JGRD..11810414C}. On a tidally-locked planet, we expect that reducing the efficiency of moist atmospheric heat transport will cool the night-side, and so increase the day-night temperature contrast. Even though global mean surface temperatures may be reduced, near the sub-stellar point we expect that a continent will have greater surface temperature than an ocean surface due to reduced cloud coverage and reduced evaporative cooling \citep[given our assumption of an `Earth-like' surface pressure, as in][]{2017A&A...601A.120B}. \citet{2010GeoRL..3718811S} demonstrate that the temperature/pressure field on a tidally-locked planet is set by an atmospheric wave response to longitudinally asymmetric surface heating. For planets large enough to contain a planetary scale Rossby wave, this can induce equatorial superrotation \citep{2011ApJ...738...71S,2013A&A...554A..69L}, which can dominate the circulation on tidally-locked planets. The location and extent of any land surface on the day-side will modify the location and amplitude of the sub-stellar surface heat flux, and so may alter the large-scale circulation of the atmosphere.

To conduct our investigation into continent-climate interaction on a tidally-locked planet,  we run climate simulations using the Met Office Unified Model, a three-dimensional atmospheric General Circulation Model, coupled to a land surface model, complete with fully interactive hydrology, and a single-layer slab ocean model. In Section \ref{sec:model} we describe our model set-up. Our results are presented in Section \ref{sec:land}. In Section \ref{sec:globalresponse} we investigate the response of primary climate diagnostics to a box-continent centered at the sub-stellar point, where we discover the introduction of a sub-stellar continent generally results in a cooler climate. In Section \ref{sec:daynight} we consider the influence of sub-stellar land on heat redistribution and day-night temperature contrasts. We find that continents can impede the ability of the atmosphere to redistribute heat to the night-side, resulting in increased day-night temperature contrasts. In Section \ref{sec:circulation}, we investigate the effect of sub-stellar land on the large-scale circulation. For continents offset to the east of the substellar point, we observe a regime change in the atmospheric circulation, namely a weaker superrotating jet and the appearance of two counterrotating mid-latitude jets. Discussion is presented in Section \ref{sec:discuss}, where we compare our work to other studies that have investigated heat redistribution on tidally-locked planets \citep[e.g.][]{2014ApJ...784..155Y,2015ApJ...806..180W}. We also discuss the implications of our findings for the potential habitability of tidally-locked terrestrial exoplanets, and make comments regarding the observability of a sub-stellar land mass, relevant to future work within the field. Finally, our conclusions are summarized in Section \ref{sec:conclusions}.

\section{Model Framework}\label{sec:model}

\subsection{General Circulation Model} \label{sec:method}

We make use of the Global Atmosphere 7.0 \citep{Walters2017} configuration of the Met Office Unified Model (UM) to simulate the atmosphere. The UM is a 3D atmospheric General Circulation Model (GCM) that solves the fully compressible, deep-atmosphere, non-hydrostatic, Navier-Stokes equations using a semi-implicit, semi-Lagrangian approach.

Sub-grid scale processes are parametrized as follows; boundary layer turbulence, including non-local transport of heat and momentum, follows \citet{2000MWRv..128.3187L} and \citet{2008BoLMe.128..117B}; cumulus convection uses a mass-flux approach based on \citet{1990MWRv..118.1483G} that includes re-evaporation of falling precipitation; multi-phase $\text{H}_{2}\text{O}$ cloud condensate and fraction amounts are treated prognostically following \citet{2008QJRMS.134.2093W} and ice and liquid precipitation formation is based on \citet{1999QJRMS.125.1607W} and \citet{2014MWRv..142.1655B} respectively. Radiative transfer is handled by the SOCRATES\footnote{To calculate radiative heating rates we use the Suite of Community Radiative Transfer codes based on Edwards and Slingo (SOCRATES), available at https://code.metoffice.gov.uk/trac/socrates} scheme \citep[described in][Section 2.3]{Walters2017} which makes use of 6 ``shortwave" bands (0.2 - 10$\,\mu$m) to treat incoming stellar radiation, and 9 ``longwave" bands (3.3$\,\mu$m - 10\,mm) for thermal emission from the planet. A correlated-\emph{k} technique is applied. All schemes are considerably improved from their original documentation, and the current incarnations are summarized in \citet{2017GMD....10.1487W}, \citet{Walters2017}, and references therein. 

\begin{deluxetable}{l|l}
\tablecaption{Orbital, planetary and atmospheric parameters used in this study.
\label{tab:params}}
\tablehead{
\colhead{Parameter} & \colhead{Value} 
}
\startdata
\underline{Orbital} & \\
Semi-major axis (AU) & 0.0485 \\
Stellar irradiance, $S$ (W m$^{-2}$) & 881.7 \\
Orbital period (Earth days) & 11.186 \\
Rotation rate, $\Omega$ (rad s$^{-1}$) & $6.501\times10^{-6}$ \\
Eccentricity & 0.0 \\
Obliquity & 0.0 \\
\cline{1-2}
\underline{Planetary} & \\
$r_{\text{p}}$ (km) & 7160 \\
$g$ (m s$^{-2}$) & 10.9 \\
\cline{1-2}
\underline{Atmospheric} \emph{($\text{N}_{2}$-dominated)} &  \\
$R$ (J kg$^{-1}$ K$^{-1}$)& 297.0 \\
$c_{\text{p}} $ (J kg$^{-1}$ K$^{-1}$) & 1039.0 \\
CO$_{2}$ Mass mixing ratio (kg kg$^{-1}$) & $5.941\times10^{-4}$ \\
Mean surface pressure, $p_{0}$ (Pa) & $10^{5}$ \\
\enddata
\end{deluxetable}

The UM has been adapted to simulate a wide range of planetary climates, including those of both gas giant planets \citep{2014A&A...561A...1M,2016A&A...595A..36A} and terrestrial planets \citep{2014GMD.....7.3059M,2017A&A...601A.120B}. For this work, the model is configured to simulate the climate of a tidally-locked terrestrial exoplanet. We choose to use the orbital and planetary parameters of ProC b, following \citet{2017A&A...601A.120B}, based on the best estimates provided by \citet{2016Natur.536..437A} and \citet{2016A&A...596A.112T}, as a `template'. The stellar spectrum is from BT-Sett1 \citep{2013A&A...556A..15R} with $T_{\text{eff}} = 3000\,\text{K}$, $g = 1000\,\text{m\,s}^{-2}$ and $\text{metallicity} = 0.3\,\text{dex}$, following \citet{2010A&A...519A.105S}.

We choose a model resolution of 2.5$^{\circ}$ longitude by 2$^{\circ}$ latitude, with 38 levels in the vertical extending from the surface to the top-of-atmosphere (40\,km), quadratically stretched to enhance resolution near the surface. The model timestep is $1200\,\text{s}$. Model parameters are presented in Table \ref{tab:params}.

As far as the atmosphere is concerned, our setup is identical to that used for the nitrogen-dominated ProC b simulations presented in \citet{2017A&A...601A.120B}. Whilst we keep our discussion general to any tidally-locked terrestrial exoplanet, this choice allows the reader to make an easy comparison with the results of \citet{2016A&A...596A.112T}, \citet{2017A&A...601A.120B} and \citet{2017arXiv170902051D}.

\subsection{The surface boundary condition}

The principal aim of this study is to investigate the effect of a sub-stellar continent on the climate of tidally-locked terrestrial exoplanets. To do this, we use the Joint UK Land Environment Simulator \citep[JULES,][]{2011GMD.....4..677B} to represent land covered surface, in conjunction with a single layer `slab' model based on \citet{2006JAtS...63.2548F} to represent ocean covered surface.

For the slab ocean, we choose a heat capacity of $10^{8}\,\text{J}\,\text{K}^{-1}\,\text{m}^{-2}$, which is representative of an ocean surface with a 24\,m mixed layer. This choice, which differs from that made in \citet{2017A&A...601A.120B} (2.4\,m mixed layer), is designed to make a distinction between the heat capacity of land and ocean. In our simulations, ice-free ocean has an albedo of $\alpha_{\text{min}} = 0.07$. We decide to include a simple representation of sea-ice following the `HIRHAM' parametrization presented in \citet{2007IJCli..27...81L} \citep[based on][]{1996JGR...10123401D}, which changes the surface albedo of the ocean if surface temperatures fall below a critical temperature, $T_{\text{c}} = 271\,\text{K}$. Below $T_{\text{c}}$, the ocean surface albedo is given by: \begin{equation} \label{eq:albedo}
\alpha_{\text{o}} = \alpha_{\text{max}}-\exp(-(T_{\text{c}}-T_{\text{s}})/2)\cdot(\alpha_{\text{max}}-\alpha_{\text{min}}),
\end{equation} where $T_{\text{s}}$ is surface temperature, $\alpha_{\text{max}} = 0.27$ is the maximum ice-albedo, and $\alpha_{\text{min}} = 0.07$ is the ice-free ocean albedo. We choose $\alpha_{\text{max}} = 0.27$ based on the mean bolometric albedo for ice calculated in \citet{2016A&A...596A.112T} for a planet with ProC b's incident stellar spectrum. For surface temperatures greater than $T_{\text{c}}$, $\alpha_{\text{o}}=\alpha_{\text{min}}=0.07$. Unlike in \citet{2017A&A...601A.120B}, ocean albedo is spectrally independent, to retain consistency with the simplicity of the sea-ice parametrization. 

\begin{deluxetable}{l|lll}
\tablecaption{Continent Configurations.
\label{tab:continents}}
\tablehead{
\colhead{Simulation} & \colhead{Latitude (deg)} & \colhead{Longitude (deg)} & Land Surface  \\
\colhead{} & \colhead{$(\lambda_{\text{min}},\lambda_{\text{max}})$\tablenotemark{a}} & \colhead{$(\phi_{\text{min}},\phi_{\text{max}})$\tablenotemark{a}} & \colhead{Coverage} 
}
\startdata
Homogeneous & & & \\
\emph{Aqua} & N/A & N/A & 0.0 \\
\cline{1-4}
Centred & & & \\
\emph{B1} & (-15,15) & (-28.75,28.75) & 4.6\% \\
\emph{B2} & (-19,19) & (-36.25,36.25) & 7.1\% \\ 
\emph{B3} & (-25,25) & (-46.25,46.25) & 11.6\% \\
\emph{B4} & (-29,29) &  (-56.25, 56.25) & 16.6\%  \\ 
\emph{B5} & (-35,35) & (-66.25,66.25) & 22.0\% \\
\emph{B6} & (-39,39) & (-73.75,73.75) & 26.8\%  \\
\emph{B7} & (-45,45) & (-86.25,86.25) & 34.4\% \\
\emph{B8} & (-49,49) & (-93.75,93.75) & 39.9\%  \\
\cline{1-4}
East-offset\tablenotemark{b} & & &  \\
\emph{E2} & (-19,19) & (1.25,73.75) & 7.1\%  \\
\emph{E4} & (-29,29) & (1.25,113.75) & 16.6\%  \\
\emph{E6} & (-39,39) & (1.25,148.75) & 26.8\% \\
\enddata
\tablenotetext{a}{Degrees from sub-stellar point, located at $(\lambda,\phi) = (0,0)$.}
\tablenotetext{b}{Named to correspond with sizes of `B' configurations, i.e. \emph{B2} and \emph{E2} are the same size.}
\end{deluxetable}

For each column in the UM where land is present,  JULES is used to simulate the surface boundary. There are four soil layers (labelled 1 to 4, layer 1 closest to the surface) with a thickness of 0.1, 0.25, 0.65 and 2\,m, for layers 1 to 4 respectively, between which soil water and heat fluxes are calculated. JULES operates a tile approach, with the surface of each land point subdivided into five types of vegetation and four non-vegetated surface types. For our simulations, we set all grid points to have 100\% bare soil coverage, where the soil is given a sandy composition. We initialize soil moisture to be 0.2, 0.5, 1.2 and 3.8\,kg\,m$^{-2}$ for layers 1 - 4 respectively. The maximum soil moisture that can be reached is defined by the volumetric soil moisture content at saturation, which is set to 0.363 based on Earth-like parameters for soil with a sandy composition. The maximum soil water content in a given layer is therefore $\rho_{\text{H}2\text{O}}\cdot0.363\cdot h_{\text{l}}\,\text{kg\,m}^{-2}$ where $\rho_{\text{H}2\text{O}}$ is the density of water and $h_{\text{l}}$ the layer depth. Evaporation from the soil is permitted; transpiration is not as there is no vegetation. Excess water is removed as surface run-off and is assumed to return to the ocean. For simplicity, land is assumed to be flat, with constant sea-level altitude. Therefore, whilst the roughness length ($10^{-3}\,\text{m}$) is higher than that of the ocean surface, and will exert an additional drag force on the near-surface flow, we do not introduce large perturbations to the flow associated with gravity wave drag, flow blocking or orographic roughness. A full description of JULES, documenting processes including surface exchange, soil fluxes and run-off, can be found in \citet{2011GMD.....4..677B}.

Without observational constraints, our choice for the land surface albedo, $\alpha_{\text{ls}}$, is informed by values found for present-day Earth. On Earth, desert land has an albedo $\alpha_{\text{ls}} = 0.42$ in the 0.85 - 4\,$\mu$m wavelength range, and $\alpha_{\text{ls}} = 0.5$ in the 0.7 - 0.85\,$\mu$m range \citep{Coakley2003}. Most of the stellar radiation incident on ProC b falls within this range \citep[see][their Figure 1]{2017A&A...601A.120B}. In reality, much of Earth's land surface has a lower albedo due to the presence of vegetation. Recognising that $0.5$ is a high value to choose for land surface albedo, we choose $\alpha_{\text{ls}} = 0.4$. We note that this choice is still significantly higher than the value of $\alpha_{\text{ls}} = 0.2$ used in the `dry planet' simulation of \citet{2016A&A...596A.112T} and simulations with land in \citet{2003AsBio...3..415J}, however it is consistent with our assumption that the soil has a sandy composition.

In order to investigate the impact of sub-stellar land on the climate we run simulations for three surface `configurations'. We use an aquaplanet simulation, \emph{Aqua}, as our control. We run eight simulations where a box-continent of varying extent is centered at the sub-stellar point. These simulations are named \emph{B1,  B2, B3, B4, B5, B6, B7, B8}, hereafter \emph{B}(1-8), to convey increasing size. Additionally, we run three simulations where a continent is introduced with its western coastline located at $\phi = 1.25^{\circ}$, so that the centre of the continent is offset to the east of the sub-stellar point These simulations are named \emph{E2, E4, E6}, to correspond with sizes of `\emph{B}' configurations, i.e. \emph{B2} and \emph{E2} are the same size. Continent bounds and fractional surface coverage for each configuration are presented in Table \ref{tab:continents}.

\section{Results} \label{sec:land}

As detailed in the introduction, the results of this study are presented in three parts. In Section \ref{sec:globalresponse} we investigate the response of primary climate diagnostics such as evaporation, precipitation and surface temperature, to box continents symmetric about the sub-stellar point. In Section \ref{sec:daynight} we describe the effect of variations in water vapor availability on the planetary energy budget and heat redistribution. In Section \ref{sec:circulation} we examine the response of the large-scale circulation to sub-stellar land. In addition to considering continents symmetric about the sub-stellar point, we introduce a continent whose center is offset to the east of the sub-stellar point. 

All simulations are run for 20 Earth years, and the results presented are temporal mean values over the final five years of simulation, unless stated otherwise.

\begin{figure*}
\plotone{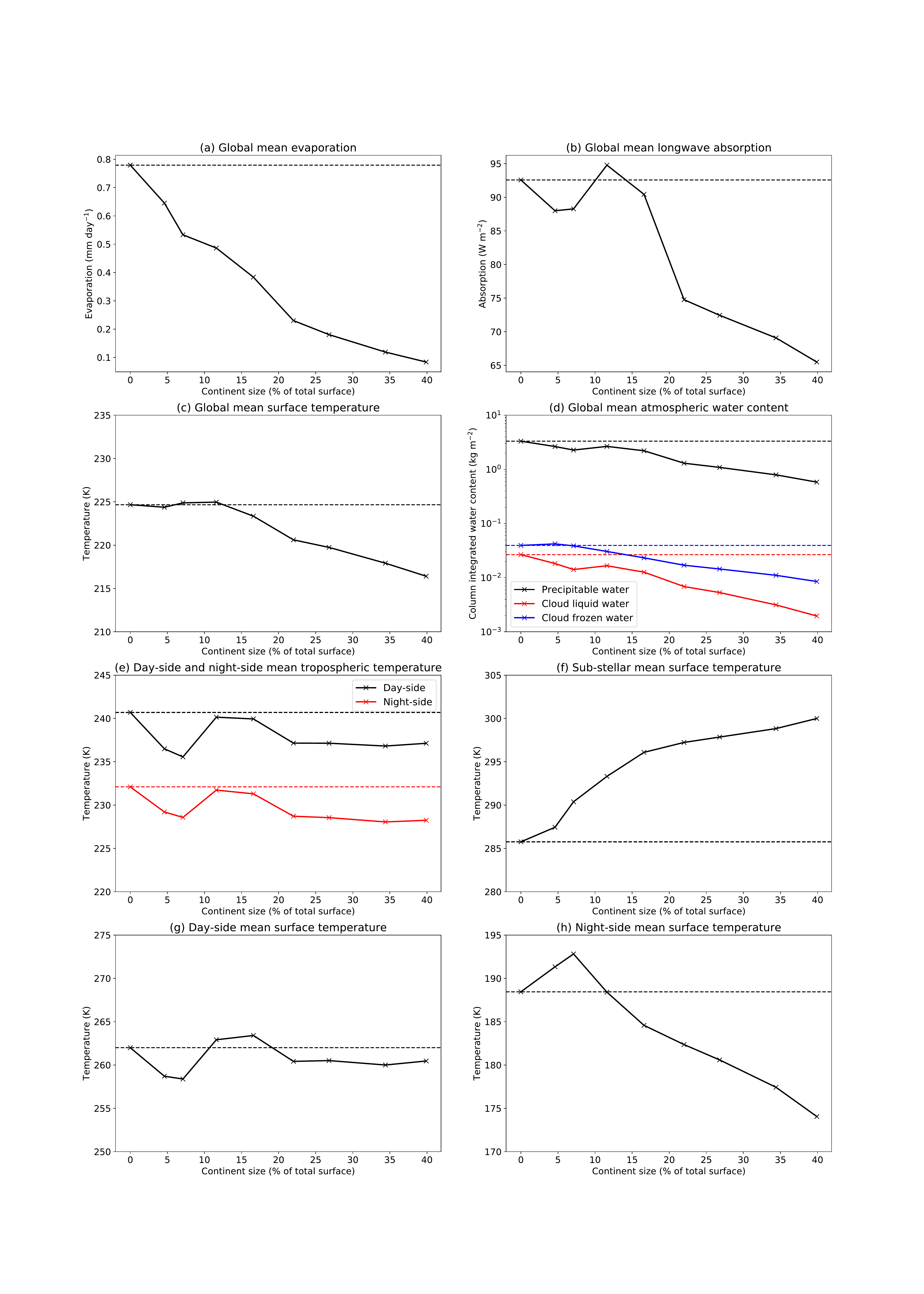}
\caption{Diagnostic quantities for the aquaplanet, \emph{Aqua}, and box continent, \emph{B}(1-8) simulations. All results presented are area-weighted spatial means over either the entire planet (global mean), the day-side, the night-side or a sub-stellar region defined by the area of the \emph{B1} continent (sub-stellar mean). In each panel, the dashed line is the value found for the \emph{Aqua} simulation.  Quantities presented are as follows: (a) global mean evaporation, (b) global mean longwave absorption, defined as $LW_{\text{abs}} = LW_{\uparrow,\text{surface}} - LW_{\uparrow,\text{TOA}}$, (c) global mean surface temperature, (d) global mean atmospheric water content, defined as $\hat{q} =\int_{0}^{\infty} \rho qdz$ for $\hat{q}\in\{\hat{q}_{\text{v}}\text{ (water vapor)},\hat{q}_{\text{cl}}\text{ (cloud liquid water)},\hat{q}_{\text{cf}}\text{ (cloud frozen water)}\}$, (e) day-side and night-side mean tropospheric temperature, defined as $T_{\text{trop}} = \left(\int^{z_{\text{trop}}}_{0}\rho Tdz\right)/\left(\int^{z_{\text{trop}}}_{0}\rho dz\right)$, where $z_{\text{trop}}=15000\,\text{m}$, (f) sub-stellar mean surface temperature, (g) day-side mean surface temperature, and (h) night-side mean surface temperature.  \label{fig:globvals}}
\end{figure*}

\subsection{Primary climate diagnostics} \label{sec:globalresponse}

Figure \ref{fig:globvals} (a) presents global mean evaporation for the aquaplanet, \emph{Aqua}, and box-continent, \emph{B}(1-8), simulations. We find that with increasing continent extent, surface evaporation decreases. The \emph{Aqua} simulation has a global mean evaporation $E = 0.78\text{\,mm\,day}^{-1}$. When the \emph{B1} continent is introduced, which covers only 5\% of the planet's surface, mean evaporation is reduced relative to the \emph{Aqua} simulation to $E = 0.65\text{\,mm\,day}^{-1}$. For the \emph{B2} continent, which covers 7\% of the total surface, mean evaporation falls further to $E = 0.53\text{mm day}^{-1}$. When land covers most of the day-side, global mean evaporation is close to zero. For example, $E = 0.08\text{\,mm\,day}^{-1}$ for the \emph{B8} simulation where roughly 75\% of the day-side surface is land. As our simulations are in equilibrium, reduced global mean evaporation is reflected by an equivalent reduction in global mean precipitation.

Reduced evaporation is associated with a reduction in atmospheric water vapor and cloud water content. In Figure \ref{fig:globvals} (d) we present global mean column-integrated water vapor, $\hat{q}_{\text{v}}$, cloud liquid water, $\hat{q}_{\text{cl}}$, and cloud frozen water, $\hat{q}_{\text{cf}}$, where $\hat{q}$ is defined as: \begin{equation} 
\hat{q} =\int_{0}^{\infty} \rho qdz, \end{equation}
where $z$ is height in m, and $\rho$ is density in $\text{kg\,m}^{-3}$, so that $\hat{q}\in\{\hat{q}_{\text{v}},\hat{q}_{\text{cl}},\hat{q}_{\text{cf}}\}$ has units of $\text{kg\,m}^{-2}$. Through comparison with \citet{2013JGRD..11810414C} we find that a sub-stellar box-continent on a tidally-locked planet has a far greater effect on atmospheric water vapor content than equatorial land on a rapidly rotating planet. \citet{2013JGRD..11810414C} find the introduction of an equatorial continent in simulations of the Archean Earth ($(\lambda_{\text{min}},\lambda_{\text{max}}) = (-38,38)$, $(\phi_{\text{min}},\phi_{\text{max}}) = (-56,56)$) results in a 17.5\% reduction in $\hat{q}_{\text{v}}$ with respect to an otherwise identical aquaplanet simulation. The continents introduced in the \emph{B4} and \emph{B5} simulations are the most comparable in size to the \citet{2013JGRD..11810414C} equatorial super-continent. These simulations see reductions in $\hat{q}_{\text{v}}$ with respect to the \emph{Aqua} simulation of 63\% and 67\%  respectively, far greater than the reduction found for the Archean Earth supercontinent.  This is because, as discussed in \citet{2017A&A...601A.120B}, evaporation on a tidally-locked planet, with which reduced $\hat{q}_{\text{v}}$ is associated, is largely restricted to a region local to the sub-stellar point where surface temperature  is great enough to permit it. To achieve a similar reduction in $\hat{q}_{\text{v}}$ on a rapidly rotating planet, we would need to introduce land to a significant portion of the equator (the evaporating region), which would require a significantly larger land mass. Our smallest box-continent (\emph{B1} simulation), which covers just 4.6\% of the surface, sees a similar reduction (20\%) in $\hat{q}_{\text{v}}$ to the \citet{2013JGRD..11810414C} super-continent that has a surface coverage roughly four times greater.

\begin{figure*}
\epsscale{1.1}
\plottwo{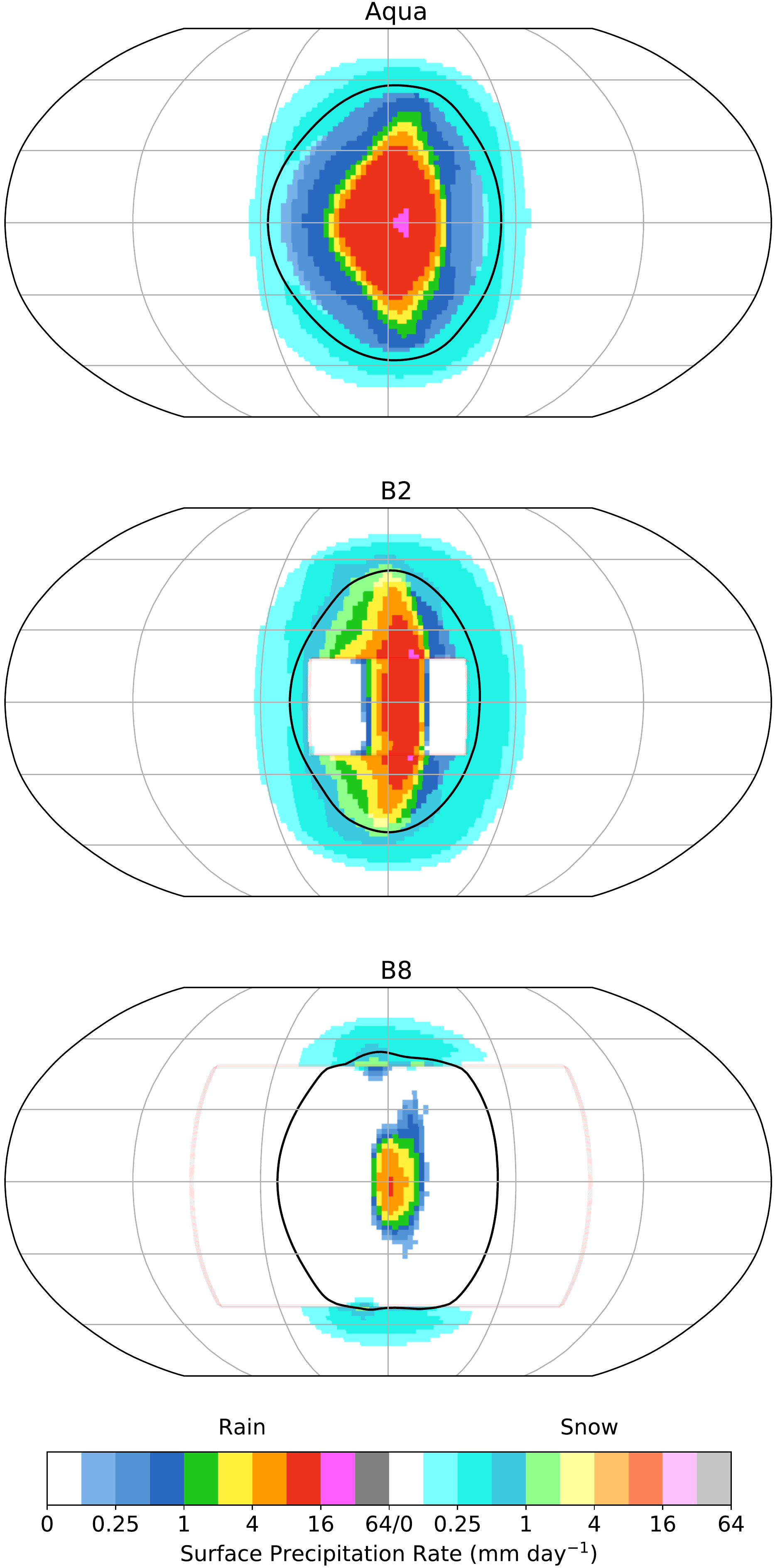}{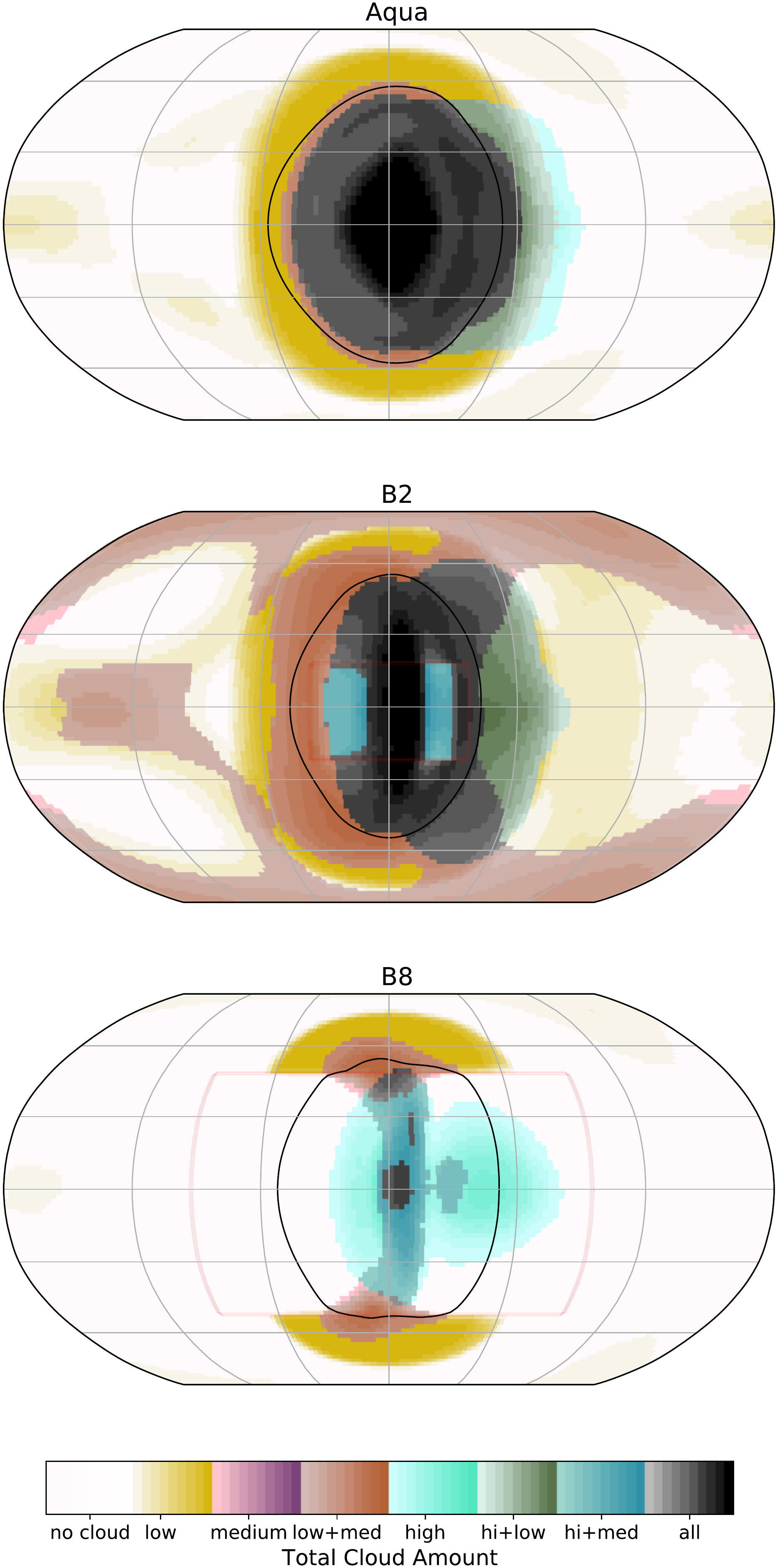}
\caption{Precipitation and cloud coverage for the \emph{Aqua} (top row),  \emph{B2} (middle row), and \emph{B8} (bottom row) simulations. Left-hand column: Precipitation. Note the non-linear scale. Right-hand column: Total cloud amount. Within each altitude range (low $<$ 2 km $<$ medium $<$ 5.5 km $< $ high) cloud coverage is given by the maximum cloud fraction (0-1) on any model level, with a contour interval of 0.1. If more that one type of cloud is present, the colorbar shows the average of the two or three cloud types present. Both columns: The black line is the mean 273.15 K contour, the red line is the continent boundary. \label{fig:precip}}
\end{figure*}

Maps of cloud coverage and surface precipitation for the \emph{Aqua}, \emph{B2} and \emph{B8} simulations are presented in Figure \ref{fig:precip}. For simulations with a box continent, precipitation is focussed on a narrow band near the center of the continent. On a tidally-locked planet, water vapor is transported towards the sub-stellar point, where precipitation is maximal, from a surrounding evaporative ring by strong boundary layer convergence. Once near the sub-stellar point, strong surface heating induces deep convection that forces water vapor upwards resulting in it being precipitated out \citep{2017A&A...601A.120B}. We have found that this process is insensitive to the introduction of a sub-stellar continent, meaning that the center of a continent can be very wet, as rain always preferentially falls near the sub-stellar point. This is important as continental silicate weathering requires precipitation to fall over land. 

We note that on a rapidly, non-synchronously, rotating planet such as Earth, where there is no preferred longitude for stellar heating and so no preferred longitude for convection and precipitation, the distribution of precipitation is not insensitive to the introduction of a land surface. This is in contrast with our tidally-locked simulations. On Earth, high land surface temperatures lead to low relative humidity (RH). Tropical rain preferentially falls over regions of highest RH \citep{2017JCli...30.4527L}, which can result in large equatorial continents with low RH becoming desert regions which see minimal rainfall \citep{2013JGRD..11810414C}. Reductions in rainfall over continental regions are then compensated for by increases elsewhere, such as over ocean regions. 

The eastern and western flanks of a sub-stellar continent, which fall outside of the precipitating band, are dry desert-like regions, which see precipitation of less than $0.125\text{\,mm\,day}^{-1}$. As continent extent increases, precipitation reduces in intensity and becomes increasingly focussed on the sub-stellar point, significantly increasing the extent of the desert regions and reducing the extent of cloud coverage overhead. We note that in the \emph{B1} and \emph{B2} simulations, night-side cloud coverage is increased with respect to the \emph{Aqua} simulation. This is associated with an increase in night-side specific humidity which occurs in spite of reduced global mean specific humidity. 

Global mean surface temperatures are presented in Figure \ref{fig:globvals} (c). Generally, the presence of a continent implies a reduction in global mean surface temperatures. The simulation with the largest box continent, \emph{B8}, has a mean surface temperature $8\,\text{K}$ lower than the \emph{Aqua} simulation. Despite globally cooler temperatures, a `habitable region' (where $T  > 273.15\,\text{K}$) is always retained. In all of our simulations $\approx 20\%$ of the surface area meets this criterion, with a maximum of 21\% for the aquaplanet simulation, and minimum values of 17\% and 18\% for the \emph{B1} and \emph{B2} simulations, respectively. We find that global mean surface temperature falls due to a reduction in longwave ($LW$) absorption by the atmosphere (i.e. the greenhouse effect), $LW_{\text{abs}} = LW_{\uparrow,\text{surface}} - LW_{\uparrow,\text{TOA}}$, due to reductions in both atmospheric water vapor and cloud coverage (TOA denotes top-of-atmosphere). $LW_{\text{abs}}$ is presented in Figure \ref{fig:globvals} (b). Increased land surface albedo ($\alpha_{\text{ls}} = 0.4$, $\alpha_{\text{o}} = 0.07$) does little to cool the planet, as it is offset by a reduction in the albedo of the atmosphere due to reduced cloud coverage. Planetary albedo remains roughly constant ($0.33\pm0.01$ for all simulations). Reduced cloud coverage serves to increase continental surface temperatures, as less radiation is absorbed and reflected in the atmosphere allowing more to be absorbed at the surface. This means that, in contrast with reduced global mean surface temperatures, surface temperatures local to the sub-stellar point increase when compared to the aquaplanet simulation. To measure the trend in sub-stellar surface temperature, in Figure \ref{fig:globvals} (f) we present temperatures averaged over the area of the \emph{B1} continent.

We note that for simulations with smaller continents (surface coverage $< 15\%$) there is little change in global mean surface temperature. However, once continent extent increases further, surface temperature falls monotonically. To understand variation in global mean temperature it is useful to consider day-side and night-side mean temperatures separately.

\subsection{Heat redistribution and day-night contrasts} \label{sec:daynight}

Figure \ref{fig:globvals} (e,g,h) presents surface and tropospheric temperatures for both the day-side and the night-side. Here tropospheric temperature is defined as the mass-weighted average temperature in the troposphere: \begin{equation} \label{eq:Ttrop}
T_{\text{trop}} = \frac{\int^{z_{\text{trop}}}_{0}\rho Tdz}{\int^{z_{\text{trop}}}_{0}\rho dz}, \end{equation} 
where $z_{\text{trop}} = 15000 \text{\,m}$ is the tropopause height and $T$ is temperature. We find that day-side surface temperatures for simulations with land remain roughly unchanged from the aquaplanet case. There is strong coupling between day-side surface and tropospheric temperature resulting from the efficient maintenance of a moist adiabat by convection.  This means that, similar to day-side surface temperatures, day-side tropospheric temperature, $T_{\text{trop,ds}}$, exhibits little deviation from the aquaplanet case upon introduction of a continent. On the night-side, whilst variation $T_{\text{trop,ns}}$ in is minimal and closely follows $T_{\text{trop,ds}}$, surface temperatures exhibit larger variation and generally fall. There is a $13\,\text{K}$ drop in mean night-side surface temperature from the \emph{Aqua} simulation to the \emph{B8} simulation. It is apparent that the variation in global mean surface temperatures is dominated by changes in night-side surface temperature. 

\begin{figure}
\plotone{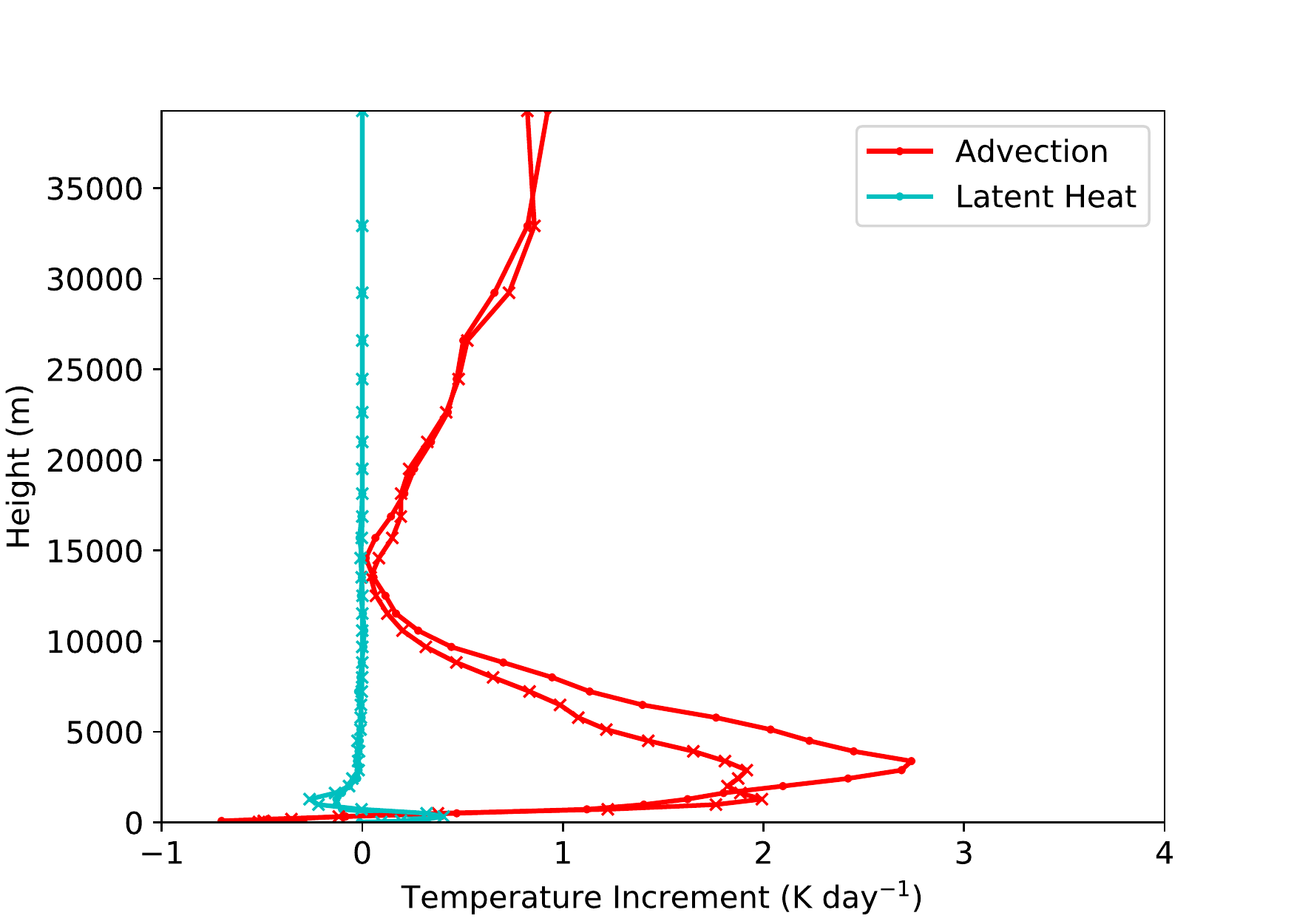}
\caption{Heat transport to the night-side for the \emph{Aqua} (dotted) and \emph{B8} (crossed) simulations. Temperature increments are presented for advection (red lines) and latent heating (blue lines). Data presented is averaged over the longitudinal and latitudinal extent of the \emph{B1} continent, centred at the anti-stellar point.} \label{fig:ns-heating}
\end{figure}

The close-coupling between day-side and night-side tropospheric temperatures must be maintained by the rapid transport of heat from the day-side to the night-side. To investigate this, we consider radiative and advective timescales. We define a radiative timescale, $\tau_{\text{rad}}$: \begin{equation} \label{eq:taurad}
\tau_{\text{rad}} = \frac{c_{\text{p}}p}{g\sigma T^{3}},
\end{equation} 
following \citet{1989artb.book.....G} and \citet{2013cctp.book..277S}. $p = 300\,\text{hPa}$ is the pressure at the jet-height, $T = 230\,\text{K}$ is a temperature typical of the night-side troposphere (see Figure \ref{fig:globvals} e), $g$ is acceleration due to gravity, and $c_{\text{p}}$ is atmospheric specific heat capacity at constant pressure (see Table \ref{tab:params}). If we additionally define an advective timescale, $\tau_{\text{adv}}$: \begin{equation} \label{eq:tauadv}
\tau_{\text{adv}} = \frac{\pi(r_{\text{p}}+h_{\text{jet}})}{U}, 
\end{equation}
where $r_{\text{p}}$ is the planetary radius (see Table \ref{tab:params}), $h_{\text{jet}} = 8000\,\text{m}$ is the height of the equatorial jet, and $U = 30\,\text{m\,s}^{-1}$ is taken as the jet-speed, then comparison between the two timescales yields $\tau_{\text{rad}}/\tau_{\text{adv}} = 5.52 > 1$. This means that heat is transported to the night-side faster than it is radiated away to space, and a weak day-night atmospheric temperature gradient is maintained by advection. We note that latent heat transport to the night-side is negligible for all simulations. Night-side advective and latent heating temperature increments are presented in Figure \ref{fig:ns-heating}.

On the night-side of the planet, there is no incident stellar radiation and so the only heat source is atmospheric heat transport from the day-side. The night-side surface receives a portion of this as radiation from the night-side atmosphere. Energy balance, therefore, dictates that night-side surface temperatures are given by: 
\begin{equation} \label{eq:nsT}
\epsilon_{\text{s}}\sigma T^{4}_{\text{ns}} = \epsilon_{\text{na}}\sigma T^{4}_{\text{na}}+F_{\text{n,turb}},
\end{equation}
where $\sigma$ is the Stefan-Boltzmann constant, $T_{\text{ns}}$ is night-side surface temperature, $\epsilon_{s}$ is surface emissivity and is a model parameter (0.985 for ocean, 0.9 for land), $T_{\text{na}}$ is night-side atmospheric temperature, $\epsilon_{\text{na}}$ is night-side atmospheric emissivity, a quantity that describes the atmosphere's ability to radiate heat, and $F_{\text{n,turb}}$ are turbulent latent and sensible heat fluxes, which remain small in all of our simulations ($F_{\text{n,turb}} = 1.4\pm0.6\,\text{W\,m}^{-2}$) when compared to radiative fluxes ($\epsilon_{\text{s}}\sigma T^{4}_{\text{ns}}, \space\sigma\epsilon_{\text{na}}T^{4}_{\text{na}} \approx 70\,\text{W\,m}^{-2}$).

\begin{figure}
\plotone{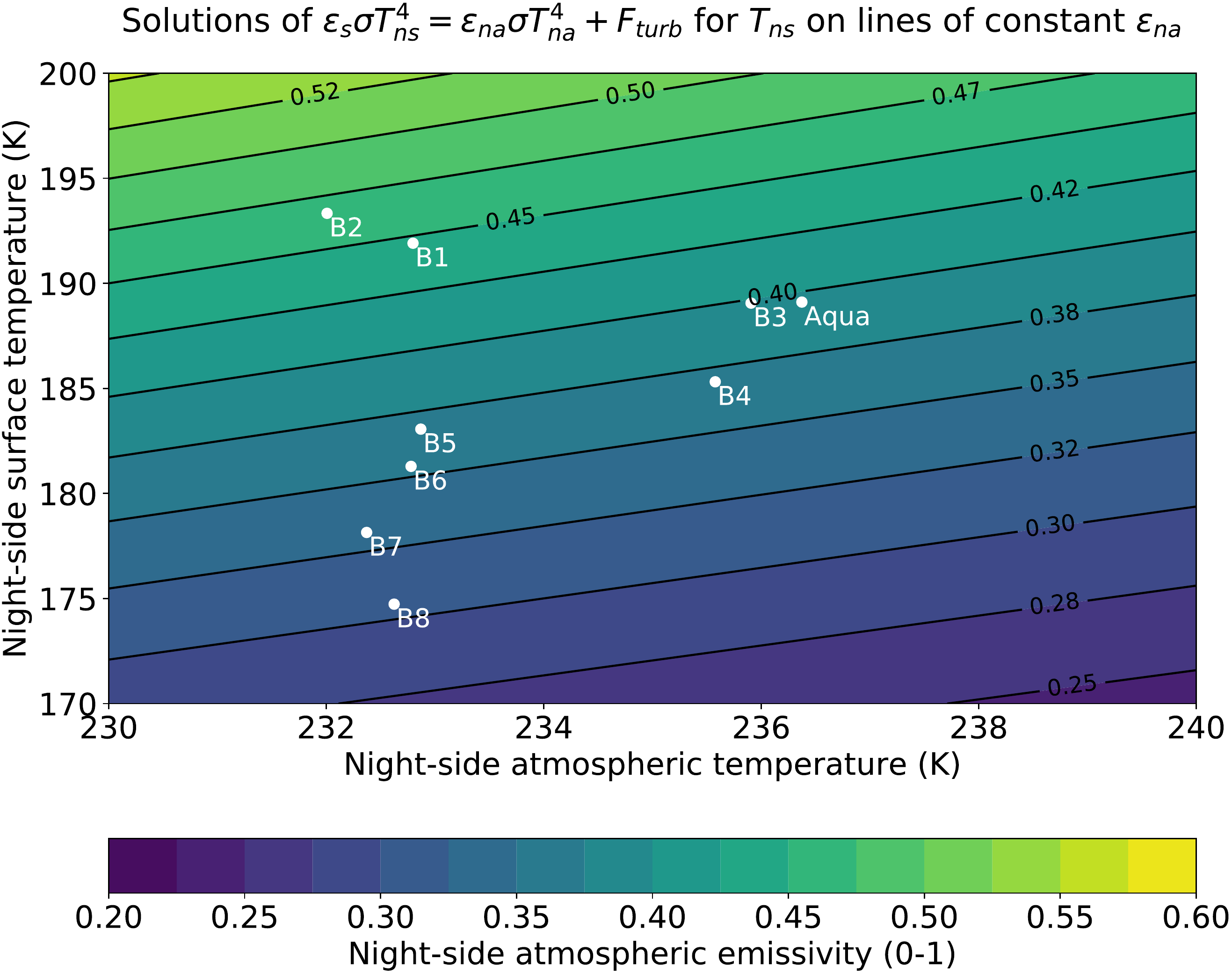}
\caption{Night-side surface temperature, $T_{\text{ns}}$ is presented on the ordinate as a function of night-side atmospheric emissivity, $\epsilon_{\text{na}}$ (colormap), and night-side atmospheric temperature (abscissa), $T_{\text{na}}$. White points overlaid are for the \emph{Aqua} and \emph{B}(1-8) simulations, where night-side mean surface temperature is used for $T_{\text{ns}}$, night-side tropospheric temperature (as defined in Equation \ref{eq:Ttrop}) is used for $T_{\text{na}}$ , and $\epsilon_{\text{na}}$ is found using Equation \ref{eq:nsT}. $\epsilon_{\text{s}}=0.985$ is used. \label{fig:eps}}
\end{figure}

\begin{figure*}
\plotone{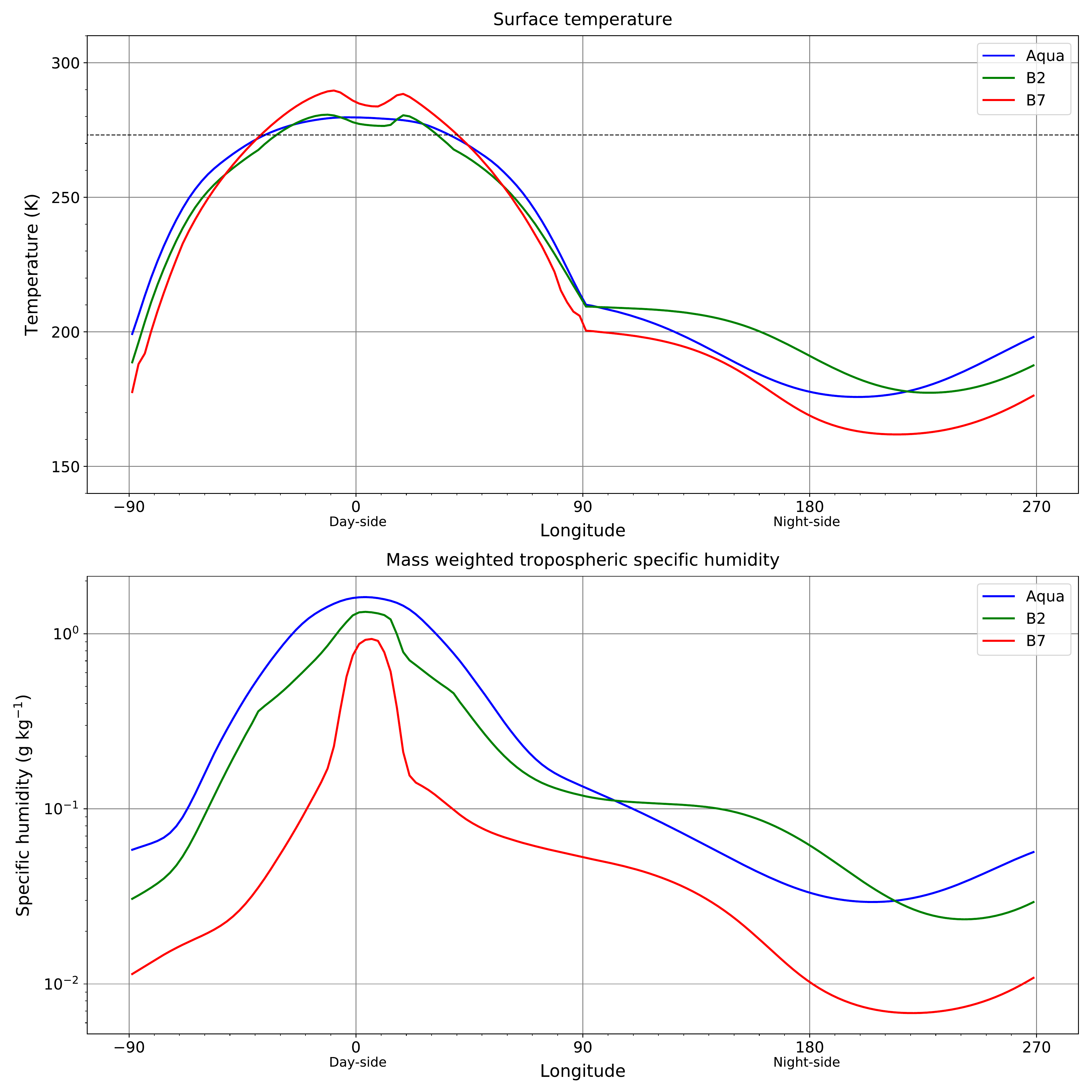}
\caption{Top panel: Meridional mean surface temperature, the dashed line is 273.15\,K. Bottom panel: Meridional mean mass weighted tropospheric specific humidity. This is the mass-weighted integral of specific humidity from the first model level to the tropopause height (15000 m).  \label{fig:merplot}}
\end{figure*}

In Figure \ref{fig:eps} we present solutions to Equation \ref{eq:nsT} for $T_{\text{ns}}$ on lines of constant $\epsilon_{\text{na}}$, where we use a constant value of $1.4\,\text{W\,m}^{-2}$ for $F_{\text{n,turb}}$. We then over-lay the results from our simulations using night-side mean surface temperature as an estimate for $T_{\text{ns}}$ and night-side mean tropospheric temperature as an estimate for $T_{\text{na}}$, and use Equation \ref{eq:nsT} to find $\epsilon_{\text{na}}$. 

Equation \ref{eq:nsT} tells us that night-side surface temperatures are dependent on night-side atmospheric temperatures and night-side emissivity. In our simulations, night-side tropospheric temperatures are strongly coupled to day-side tropospheric temperatures (see Figure \ref{fig:globvals} e). Day-side tropospheric temperatures stay roughly constant, so whilst there is some cooling in the night-side, tropospheric temperatures do not deviate far from the aquaplanet case, as a result of rapid advective heat transport. Solving Equation \ref{eq:nsT}, using $T_{\text{na}} = T_{\text{trop,ns}}$, we find $\epsilon_{\text{na}}$ is not constant for our simulations (see Figure \ref{fig:eps}). Night-side atmospheric temperature variation alone is not substantial enough to result in the surface temperature variation observed in our simulations. Instead, it is apparent that variation in $\epsilon_{\text{na}}$ is largely responsible.

Atmospheric emissivity appears to be positively related with atmospheric water content. For our simulations with `large' continents (surface fraction $> 15\%$), reduced day-side evaporation is associated with the reduced transport of moisture to the night-side. As a result the night-side atmosphere becomes drier, and is less able to absorb and radiate heat. This means that the night-side atmosphere is able to receive a smaller fraction of energy from the day-side, and a corresponding smaller fraction is passed to the night-side surface through longwave radiation from the atmosphere. Consequently, night-side surface temperatures are reduced. For the smaller \emph{B1} and \emph{B2} simulations, where the atmosphere remains relatively wet whilst maintaining increased sub-stellar surface temperatures with respect to the \emph{Aqua} simulation, more water vapor and cloud water is transported to the night-side increasing night-side emissivity. As a result, night-side surface temperatures are increased.

In Figure \ref{fig:merplot} we present meridionally meaned surface temperature, and specific humidity (SH). On the night-side there is a clear correlation between surface temperature and SH. For the larger continents investigated (\emph{B7} provides an example) we find that reduced night-side SH results in reduced surface temperatures, particularly between $180^{\circ}\le\phi\le270^{\circ}$ where the reduction in atmospheric water vapor content between the continental and aquaplanet cases is at its maximum. Meanwhile, for the \emph{B1} and \emph{B2} simulations (\emph{B2} is presented in Figure \ref{fig:merplot}), we observe that where night-side specific humidity is increased with respect to the \emph{Aqua} simulation, so is night-side surface temperature. In the case of both `larger' and `smaller' continents, variation in night-side surface temperature is dominated by variation in night-side atmospheric emissivity, $\epsilon_{\text{na}}$, caused by variation in atmospheric water vapor and cloud water content.

On the day-side, variation in surface temperature is governed by several processes. The introduction of a continent reduces evaporation and so cloud coverage overhead, which increases the radiation incident on the surface below, allowing continental surface temperatures to rise. However, reduced cloud coverage and atmospheric water vapour content also reduces the day-side greenhouse effect, allowing more radiation from the surface to escape to space.  For each of our simulations, whether the day-side temperature increases or decreases is a competition between these two effects, and the amount of heat lost to the night-side. For the majority of our experiments, the day-side surface temperature is reduced, although the change is always small.

The day-side surface energy balance is given by \begin{equation} \label{eq:dsT}
\epsilon_{\text{s}}\sigma T^{4}_{\text{ds}} = \epsilon_{\text{da}}\sigma T^{4}_{\text{da}}+F_{\text{turb,d}} + \frac{S}{2}(1-\alpha_{\text{p}}), 
\end{equation}
where the terms follow the same notation convention as used in Equation \ref{eq:nsT}, with the subscript $d$ denoting a day-side quantity. $S$ is the top of the atmosphere stellar flux, and $\alpha_{\text{p}}$ is the planetary albedo. Unlike on the night-side, the day-side surface is heated by both stellar radiation \emph{and} longwave radiation from the atmosphere. Consulting Figure \ref{fig:merplot}, on the day-side we observe that SH, and so atmospheric emissivity, has little influence on surface temperatures. To first-order, this is because the stellar radiation absorbed by the surface, $\frac{S}{2}(1-\alpha_{\text{p}})$, is comparatively much larger than the radiation received by the surface from the atmosphere. Therefore any variation in longwave radiation emitted to, and absorbed by, the surface comprises a much smaller fraction of the surface energy balance than on the night-side, where longwave radiation from the atmosphere is the only significant source of surface heating. Towards the terminators on the day-side, where $S$ provides a smaller contribution and heating from the atmosphere is more important, the drier atmosphere means that surface temperatures for simulations with continents fall below aquaplanet surface temperatures.

\begin{figure}
\plotone{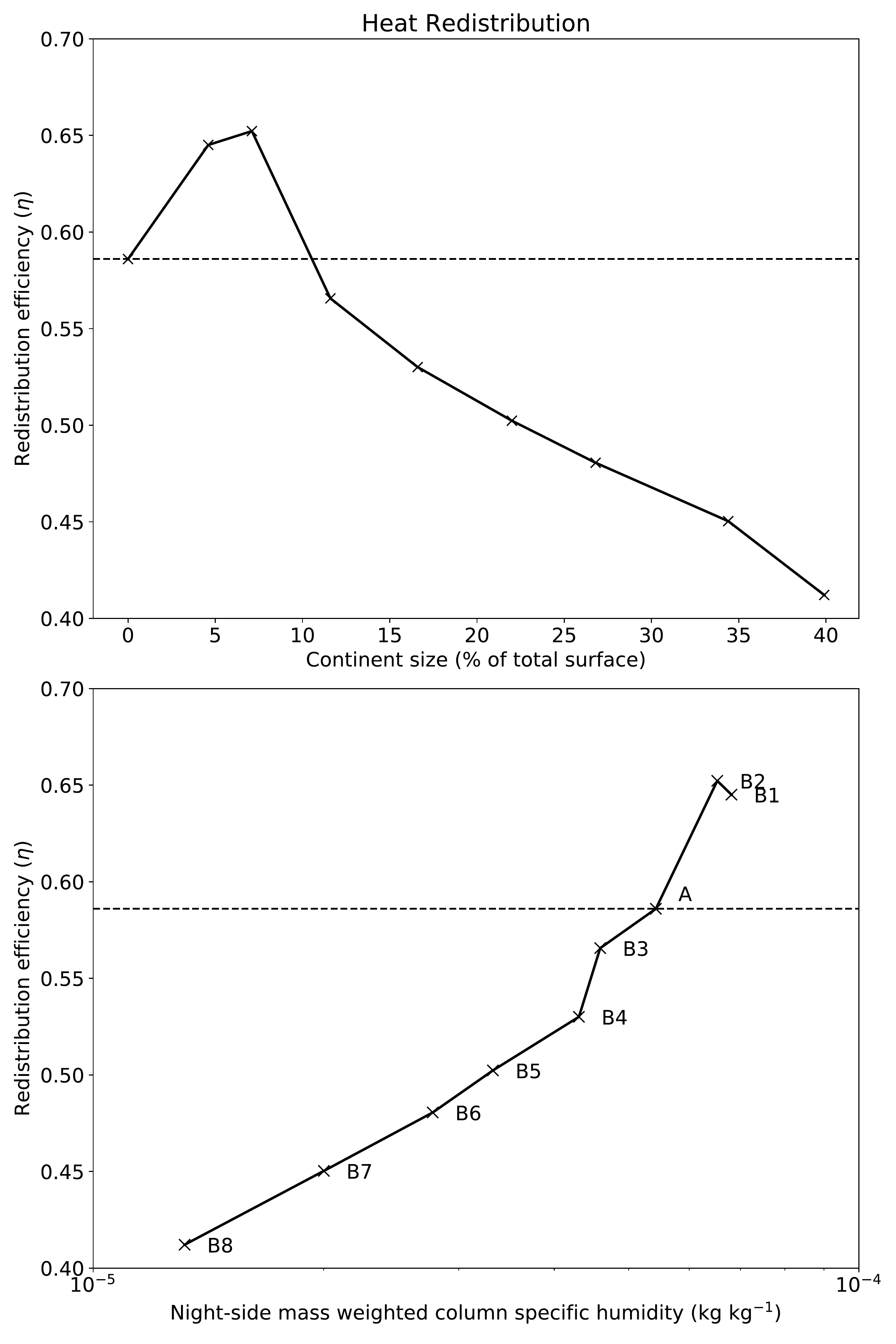}
\caption{Heat redistribution efficiency, $\eta$, defined as the ratio of mean night-side outgoing longwave radiation (OLR) over mean day-side OLR. Top panel: Heat redistribution efficiency against continent extent. Bottom panel: Heat redistribution efficiency against night-side mean tropospheric specific humidity. \label{fig:eta}}
\end{figure}

We have found that night-side atmospheric water content is important in determining the redistribution of heat from the day-side to the night-side. A useful metric to quantify this is the day-night heat redistribution efficiency, defined as the ratio of mean night-side outgoing longwave radiation (OLR) over mean day-side OLR, $\eta$, following \citet{2013A&A...554A..69L}. $\eta$ is presented in Figure \ref{fig:eta}, both as a function of continent extent (top panel) and night-side specific humidity (bottom panel). As specific humidity is reduced, so is redistribution efficiency. The one exception to this is the \emph{B2} simulation, however, this simulation has more night-side cloud than the \emph{B1} simulation. This contributes to increased night-side emissivity which results in it having a slightly higher redistribution efficiency. The two smallest continental configurations, \emph{B1} and \emph{B2}, increase the redistribution efficiency.  Increased $\eta$ for these simulations may explain why they have slightly cooler day-side surface temperatures (see Figure \ref{fig:globvals} g). In this scenario, the night-side atmosphere is better at radiating energy to space which requires that more heat is transported to it from the day-side, resulting in cooler day-side temperatures \citep{2014ApJ...784..155Y}. Similarly, we can see that reduced $\eta$ for larger continents may assist in stabilizing day-side temperatures and prevent them from deviating further from the \emph{Aqua} simulation, as less heat is transported to the night-side atmosphere. 

In spite of reduced night-side temperatures, no night-side location ever experiences temperatures less than $125\,\text{K}$, the temperature where $\text{CO}_{2}$ could condense onto the surface \citep[given the partial pressure of CO2 used in our simulations, and][]{2017A&A...601A.120B}. This means that the $\text{CO}_{2}$ greenhouse effect is retained, which helps the planet to avoid transition into a `snowball' state where surface temperatures are below $273.15$\,K everywhere.

\subsection{Large-scale circulation}\label{sec:circulation}

\begin{figure}
\plotone{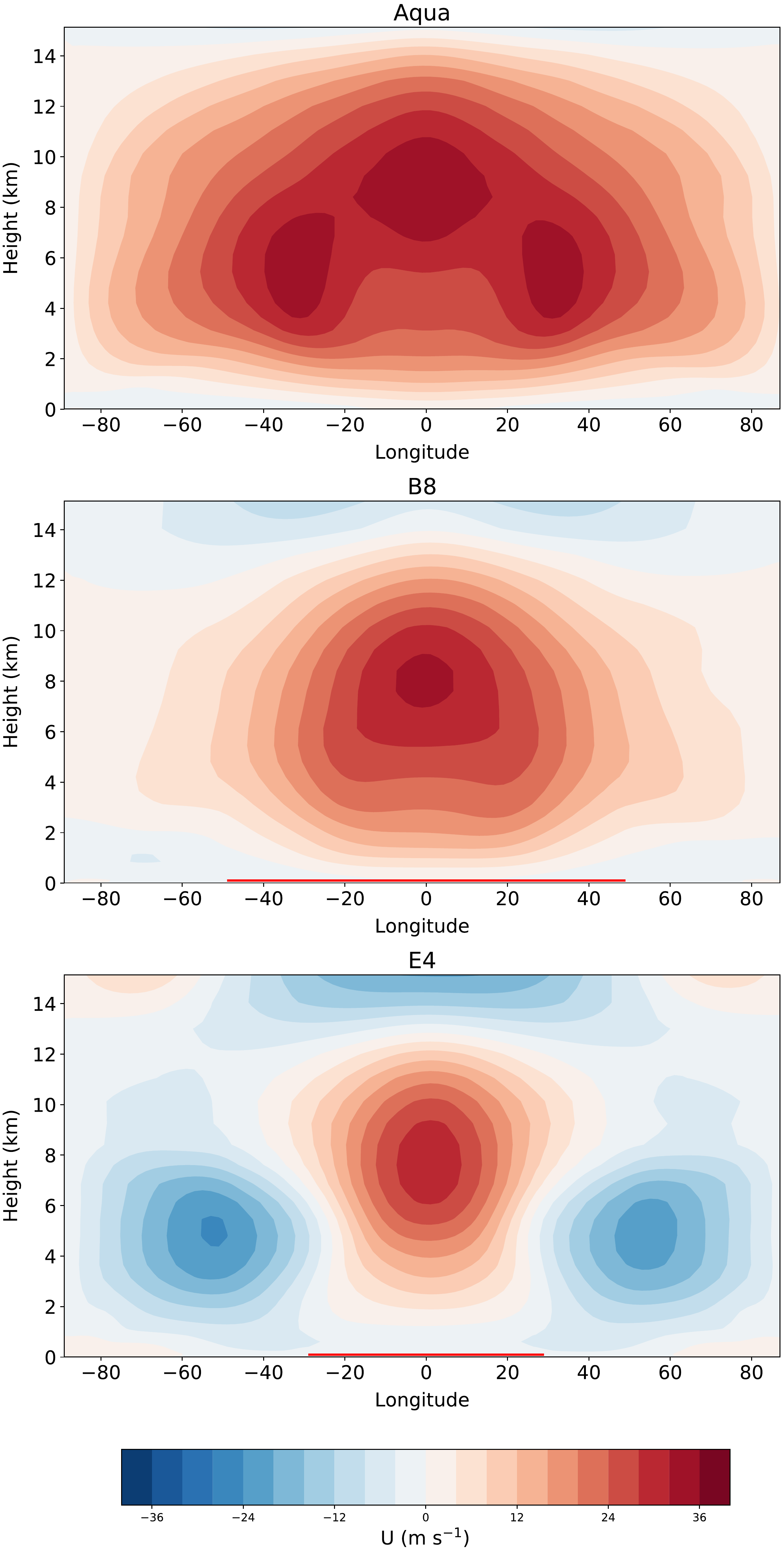}
\caption{Zonal mean zonal wind, $U$ (m s$^{-1}$). Top panel: \emph{Aqua} simulation. Middle Panel: \emph{B8} simulation (the largest box continent). Bottom panel: \emph{E4} simulation (an east-offset continent). The red line denotes the latitudinal extent of the continent.\label{fig:zonal}}
\end{figure}

Finally, we consider the effect of a sub-stellar land surface on the large-scale circulation. Zonally averaged zonal winds for the \emph{Aqua}, \emph{B8}, and \emph{E4} simulations are presented in Figure \ref{fig:zonal}. In all of our simulations, equatorial superrotation is induced. The emergence of this phenomenon can be understood by considering the dimensionless equatorial Rossby deformation length, $\mathcal{L}$: \begin{equation} \label{eq:rossby}
\mathcal{L} = \frac{L_{\text{Ro}}}{r_{\text{p}}} = \sqrt{\frac{R}{c_{\text{p}}}\frac{\sqrt{c_{\text{p}}T_{\text{e}}}}{2\Omega r_{\text{p}}}}, 
\end{equation} where $L_{\text{Ro}}$ is the equatorial Rossby deformation radius and $r_{\text{p}}$ is the planetary radius \citep{2015ApJ...806..180W,2015MNRAS.453.2412C}. $R$ is the specific gas constant for dry air, $c_{\text{p}}$ is atmospheric specific heat capacity at constant pressure and $\Omega$ is the planetary rotation rate (see Table \ref{tab:params}). $T_{\text{e}}=\left[\left(1-\alpha_{\text{p}}\right)S/4\sigma\right]^{\frac{1}{4}}$ is the effective radiative temperature of the planet. For $\mathcal{L}$ approximately equal to  (or less than) 1, a planetary scale Rossby wave can exist. For our planet, $\mathcal{L} = 1.22$, so the planet is large enough to contain a planetary scale Rossby wave \citep{2013A&A...554A..69L,2015ApJ...806..180W}. In this regime, \citet{2010GeoRL..3718811S} demonstrate that strong day-night radiative heating contrasts present on tidally-locked planets trigger the formation of standing, planetary-scale equatorial Rossby and Kelvin waves. These are similar in form to those found in shallow-water solutions for the `Matsuno-Gill' model \citep{Mat66, 1980QJRMS.106..447G}. The solutions show that when a longitudinally asymmetric heating is applied at the equator, equatorial Kelvin waves exhibit group propagation away from the heating to the east, Rossby waves located polewards of the heating exhibit group propagation away to the west. As these waves propagate away from the heating, they transport energy. \citet{2011ApJ...738...71S} studies the interaction of these waves with the mean-flow using a hierarchy of one-layer shallow water models and 3D GCM simulations of hot Jovian planets. They find that the latitudinally varying phase shift induced by the alternate propagation of the Rossby and Kelvin waves `tilts' the wind vectors northwest-to-southeast in the northern hemisphere, and southwest-to-northeast in the southern hemisphere. This serves to pump eastward momentum to the equator inducing superrotation. 

\begin{figure*}
\plotone{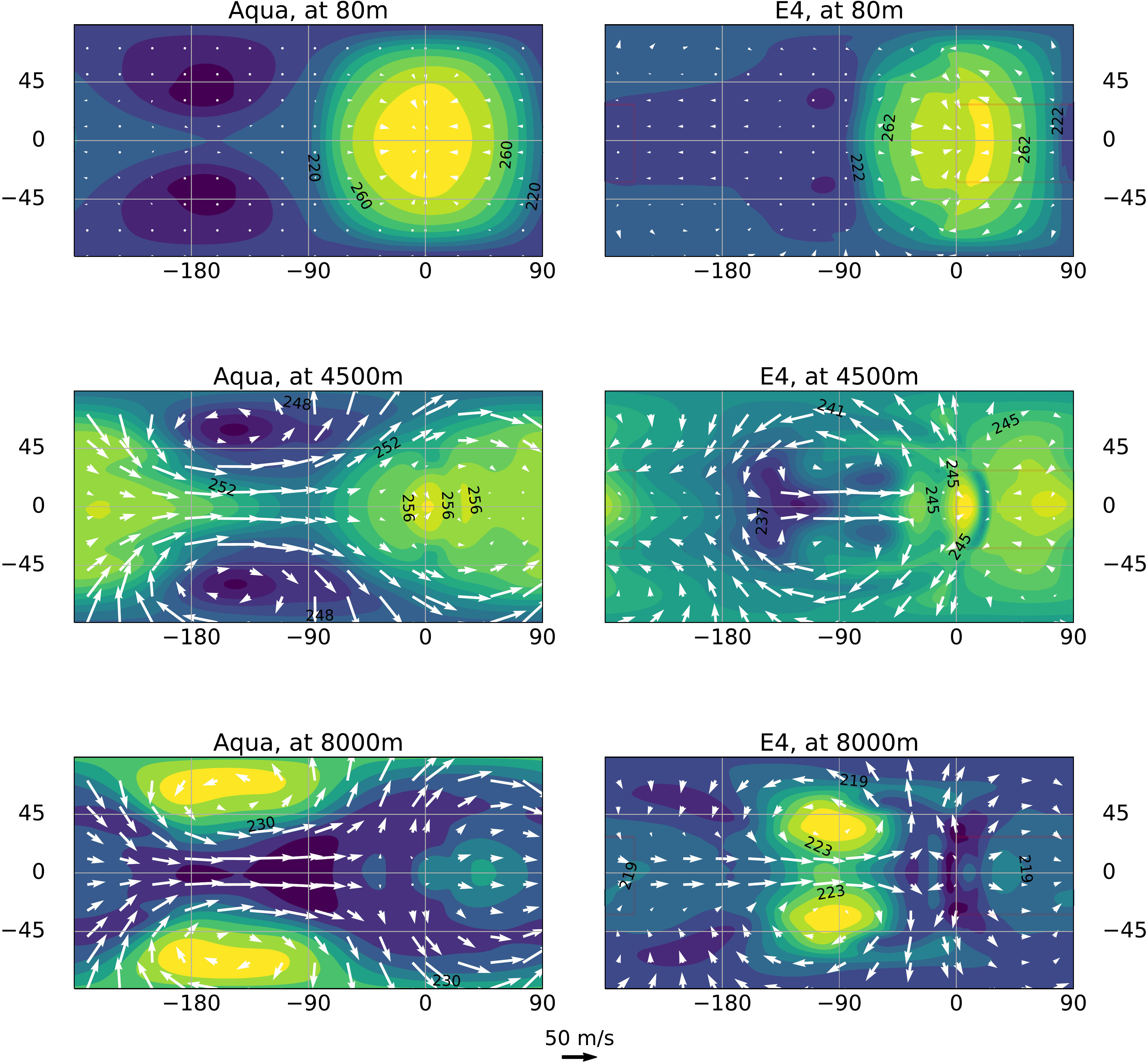}
\caption{Contours are the atmospheric temperature at the height specified in the relevant figure title. For the 80 m Figures, the contour interval is 10 K. For the 4500 m and 8000 m panels, the contour interval is 1 K. Wind vectors are presented as white arrows, and share the same normalization across panels (See the 50 ms$^{-1}$ arrow). Note: the color scales are individual to each panel, and are intended to show spatial \emph{variation} in temperature. The red line denotes continent extent for the \emph{E4} figures. \label{fig:slice}}
\end{figure*}

For increasing continent extent, the thermal forcing at the substellar point is increased as a result of increased surface temperatures over land, thus increasing the forcing amplitude. \citet{2011ApJ...738...71S} find that increased forcing amplitude should increase the strength of equatorial superrotation. We find that with increasing continent extent, the jet becomes more `focussed' on the equator. This is observed in \citet{2013A&A...554A..69L} when the equatorial jet \emph{increases} in strength. However, in our simulations, zonal wind speeds \emph{decrease} with increased forcing amplitude (see Figure \ref{fig:zonal}). We speculate that whilst increased forcing is acting to strengthen the equatorial jet, reduced moisture convergence at the sub-stellar point weakens the strength of convection, reducing vertical wind speeds. In turn, this results in a reduction in the jet speed. To add to our discussion on heat redistribution, we note that as flow becomes increasingly focussed on the equator, the Coriolis force is likely to hamper effective heat redistribution to mid-latitudes on the night-side \citep{2013A&A...554A..69L}. This may be contributing to reductions in $\eta$ found in our simulations.

In addition to considering box-continents centred at the sub-stellar point, we also consult the results of simulations \emph{E2, E4} and \emph{E6}, where a box-continent is offset to the east of the sub-stellar point. This serves to `shift' the surface thermal forcing maximum eastwards. For the \emph{E2} simulation, the large-scale circulation takes the same form as in the \emph{Aqua} and \emph{B} cases; namely, there is a superrotating jet at the equator. For the \emph{E4} and \emph{E6} simulations we observe regime change. The equatorial jet is substantially weakened and two counterrotating mid-latitude jets are introduced (see Figure \ref{fig:zonal}, the \emph{E4} simulation). 

Atmospheric temperature and wind vectors for the \emph{Aqua} and \emph{E4} runs are presented in Figure \ref{fig:slice},  where for both of the simulations we can see the temperature field somewhat resembles the form of the Matsuno-Gill standing wave response. This is indicated by the presence of two off-equatorial Rossby nodes found to the west of the sub-stellar point, and an equatorial Kelvin node to the east of the sub-stellar point. The \emph{Aqua} simulation exhibits a superrotating jet that benefits from momentum convergence from the mid-latitudes, as proposed in \citet{2011ApJ...738...71S}. At lower altitudes, the \emph{E4} and \emph{E6} simulations do not see eastward momentum imported to the equator from the mid-latitudes. A similar momentum `pumping' mechanism is observed, however it is reversed. Wind vectors are now tilted northeast-to-southwest in the northern hemisphere, and southeast-to-northwest in the southern hemisphere, meaning that westward momentum is converged towards the equator. As a result, the equatorial jet is reduced in strength and latitudinal extent. 

By introducing the continent offset to the east of the sub-stellar point, we `shift' the surface heat source to the east. This is the case as whilst the latent heating associated with deep convection still occurs over ocean regions to the west of the sub-stellar point, lower altitude turbulent boundary layer heating, communicated via dry convection, is introduced over the continent to the east of the sub-stellar point. If the surface heating were the only heat source for the atmosphere, we expect this would only serve to shift the entire pattern by the same distance, in such a way that its structure remained the same. Instead, we observe that the Rossby gyres move closer to the sub-stellar point and decrease in extent, and that the structure of the temperature field surrounding the Kelvin node is altered so that isotherms are now orientated in such a way that they direct eastward momentum away from the equator (see Figure \ref{fig:slice}, particularly in the 4500\,m panel). We suspect this arises because the surface is not the only source of heating for the atmosphere. The atmosphere is also directly heated by incident shortwave radiation\footnote{Shortwave heating of the atmosphere is increased for planets orbiting M-type stars, as the stellar spectrum is shifted towards the longwave, making it easier for water vapour to absorb the incident radiation \citep{2017A&A...601A.120B}.} and latent heating from deep convection in the upper troposphere \citep[see][their Figure 4]{2017A&A...601A.120B}. By offsetting the surface heating to the east, we take the `surface' and `direct' heating sources `out of alignment'. This implies less resonance between atmospheric waves at different altitudes, weakening the Kelvin and Rossby wave response. As a result, the temperature field is altered and geostrophic balance now requires that flow is `tilted' in such a way that it supplies westward momentum to the equator. Evidence for the cause being longitudinally offset surface and direct stellar heating can be observed by noting that the 8000\,m panels for the \emph{Aqua} and \emph{E4} simulations, where equatorial superrotation is still observed for the \emph{E4} simulation, display more similarity than the 4500\,m panels (Figure \ref{fig:slice}), and that the mid-latitude counterrotating jets are most prominent at lower altitude (Figure \ref{fig:zonal}). This is because further from the surface, offset surface heating has less of an effect, and the circulation is more similar to the regime where the heating sources are in alignment. 

A consequence of this regime change is increased night-side atmospheric water content, which allows increased heat redistribution to the night-side. For the \emph{Aqua} simulation we have $\eta = 0.57$, while for the \emph{E4} simulation $\eta = 0.65$.  As a result, we see an increase in night-side mean temperatures ($T_{\text{ns, \emph{Aqua}}} = 188.4\,\text{K}$, $T_{\text{ns, \emph{E4}}} = 197.2\,\text{K}$) and a decrease in day-side mean temperatures ($T_{\text{ds, \emph{Aqua}}} = 262.0\,\text{K}$, $T_{\text{ds, \emph{E4}}} = 257.6\,\text{K}$). Overall there is a slight ($\approx2$\,K) increase in global mean temperatures for the \emph{E4} simulation, compared to the \emph{Aqua} simulation.

\section{Discussion} \label{sec:discuss}

\subsection{Heat transport}  
A thorough investigation of atmospheric heat redistribution on tidally-locked planets is presented in  \citet{2014ApJ...784..155Y}, where a two-column model is applied to investigate the effects of water vapor and clouds on day-night contrasts in thermal emission. Varying day-side emissivity, $\epsilon_{\text{da}}$, with fixed night-side emissivity, $\epsilon_{\text{na}} = 0.5$, they find that the magnitude of variation in $T_{\text{ds}}$ is relatively small, with just a $\approx 7$\,K increase for $\epsilon_{\text{da}} = 0 \rightarrow \epsilon_{\text{da}} = 1$. Performing the same experiment but for fixed $\epsilon_{\text{da}} = 0.5$ and varying $\epsilon_{\text{na}}$ between zero and 1, they find reducing $\epsilon_{\text{na}}$ substantially warms the day-side surface with temperature, $T_{\text{ds}}$, increasing by $\approx 45$\,K as $\epsilon_{\text{na}}$ is reduced from 1 to zero. Further, they report that global mean temperatures decrease, implying a substantial decrease in night-side surface temperature, $T_{\text{ns}}$. The sensitivity to night-side emissivity is explained via an analogy whereby the night-side is described to behave as a ``radiator fin", similar to that presented in \citet{1995JAtS...52.1784P} for the Earth's tropics, that is found to be able to radiate to space easily. When $\epsilon_{\text{na}}$ is reduced, night-side heat loss to space is reduced which \emph{necessarily requires} a reduction in heat transport to the night-side \citep{2014ApJ...784..155Y}. Similarly, when $\epsilon_{\text{na}}$ is increased, an increase in heat transport from the day-side to the night-side is \emph{required}. Our results clearly show the theory presented in \citet{2014ApJ...784..155Y} in operation in a 3D GCM. Furthermore, our findings, and those of \citet{2014ApJ...784..155Y}, demonstrate that not only does this mechanism plays an important role in setting the day-night thermal emission contrast (quantified in our results as $\eta$), but also in controlling night-side surface temperatures. This arises because, as presented in Section \ref{sec:daynight}, the night-side atmosphere is the only source of heat for the surface, and so $T_{\text{ns}}$ is directly dependent on $\epsilon_{\text{na}}$. A slight difference between our results and those of \citet{2014ApJ...784..155Y} is that for decreasing $\epsilon_{\text{na}}$ we do not see a substantial increase in day-side surface temperature. This is because in our simulations $\epsilon_{\text{da}}$ and $\epsilon_{\text{na}}$ decrease together in tandem, unlike in their two-column model where one is varied while the other is fixed. This means that on the day-side, both the greenhouse effect and the direct absorption of stellar radiation by the atmosphere are reduced, thus offsetting any increase in day-side surface temperatures that would result from the reduced export of heat to the night-side. 
 
We have studied a climate regime where equatorial superrotation is present. As discussed in Section \ref{sec:circulation}, this is made possible as the planet is able to `contain' a planetary scale Rossby wave \citep{2013A&A...554A..69L,2015ApJ...806..180W}. This was quantified by considering the equatorial Rossby deformation length, $\mathcal{L}$  (Equation \ref{eq:rossby}). Planets where $\mathcal{L}$ is significantly greater than 1 are too small to contain a planetary Rossby wave. In this scenario, stellar-to-antistellar point circulation occurs, and no jets appear \citep{2013A&A...554A..69L}. \citet{2015ApJ...806..180W} finds in this regime that the circulation takes the form of a single planetary-sized convection cell on the day-side, with a `slower residual circulation' on the night-side. In this regime, the author suggests that the planetary boundary layer, and not the large-scale circulation, is the key term in deciphering the planetary energy balance. We expect that a reduction in day-night heat redistribution induced by the introduction of a sub-stellar continent would be observed in the stellar-antistellar circulation regime, as the reduction is largely associated with reduced day-night transport of water vapour. As there is still a circulation that connects the two hemispheres we expect that water vapour transport to the night-side would still be reduced, resulting in a lower $\epsilon_{\text{na}}$ (as we have in our simulations) which would serve to hamper the efficiency of heat redistribution in the atmosphere. To ascertain whether the effect of sub-stellar land, as a function of land fraction, would be more or less pronounced in the stellar-antistellar regime would further study, and is left for future work. 

\subsection{Habitability}
We have presented results for simulations of Proxima Centauri b with a $1\,\text{bar N}_{2}$, $5.941\times10^{-4}\,\text{kg\,kg}^{-1}\ \text{CO}_{2}$ atmosphere. For this configuration, we have found that a region where surface temperatures are above freezing is always retained. Thus the conclusion that stable surface liquid water may be present on ProC b, presented in \citet{2017A&A...601A.120B} and \citet{2016A&A...596A.112T}, is robust to the introduction of a sub-stellar land mass. Furthermore, surface temperatures never fall below $125$\,K, and so condensation of $\text{CO}_{2}$ onto the surface is avoided, and the $\text{CO}_{2}$ greenhouse effect is retained. Our results, therefore, suggest that sub-stellar land should not preclude ProC b from being a potential environment for life. This is in agreement with the `\emph{Day-land}' simulation presented in \citet{2017arXiv170902051D}. The fact that introducing a continent does not necessarily compromise habitability is important; as an exposed continental surface is required to facilitate an effective carbon-silicate weathering cycle \citep{2012ApJ...756..178A}.  

We can, however, conceive of situations where sub-stellar land could compromise the prospective habitability of a planet. For example, consider the case of a planet identical to the one in this study, but that is located further from its host star in such a way that it is cooler whilst remaining habitable. Whilst our simulations have not exhibited night-side cooling to below $125$\,K, a cooler planet located near the outer edge of the habitable zone might have minimum night-side temperatures below $125$\,K if sub-stellar land were present. For such a planet, our simulations suggest that the presence of sub-stellar land moves the outer limit of the habitable zone inwards. \citet{2013ApJ...771L..45Y} propose that clouds can provide a stabilizing feedback that expands the habitable zone at its inner boundary. They identify that as surface temperatures rise, so will the intensity of convection which will produce more thick cloud at the sub-stellar point, thus increasing the albedo and so reducing surface temperature. It is clear from our results that the presence of a large sub-stellar land mass will reduce the effectiveness of this mechanism, as cloud coverage is substantially reduced (see Figure \ref{fig:precip}), and thus makes a reduced contribution to the planetary albedo. Furthermore, whilst continental carbon-silicate weathering should balance outgassing of CO2 and increasing stellar irradiance, the maintenance of this process is vulnerable to small changes in atmospheric composition and pressure, via the `enhanced sub-stellar weathering instability' described in \citet{2011ApJ...743...41K}. \citet{2011ApJ...743...41K} find that if a decrease in day-night temperature gradient induced by an increase in atmospheric pressure, and hence temperature, requires a reduction in sub-stellar temperatures, then this in turn will reduce the weathering rate which will further increase atmospheric pressure and temperature. In this way, a tidally-locked planet with an active weathering cycle may be vulnerable to a runaway feedback where the weathering cycle can dramatically increase or decrease in efficiency; leading the planet towards either atmospheric collapse or a runaway greenhouse scenario.  \citet{2011ApJ...743...41K} suggest that such a feedback could be induced by changes in pressure resultant from volcanism or `mountain-building'.

\subsection{Future work}

Several questions have been raised by our results that require further study, beyond the scope of this work. Firstly, our model incorporates sophisticated parametrizations of convection and cloud processes. This may lead to more accurate capturing of the effects of these processes on the large-scale climate than more simple approaches \citep[e.g. the convective adjustment and diagnostic cloud schemes used in][]{2016A&A...596A.112T}. However, sophisticated treatments such as those used in this study have been developed to represent these processes accurately for Earth, and hence require more extensive study at higher resolution, to ensure the robust capturing of these processes across different planetary environments. 

We have demonstrated that the climate is sensitive to changes in the surface boundary condition. This suggests that the climate may well be sensitive to further characteristics such as land orography and friction not considered in this work, which would alter the dynamics of the atmosphere, and the surface evaporation efficiency. Additionally, there may be scope to capture the overall trends of the climate sensitivity to a land surface via more simple, faster parametrizations. As part of our early preparation for this work we found that representing a land surface by means of locally restricting surface evaporation and changing surface heat capacity yielded similar results to those retrieved using the full land-surface model. 

Of course, the inclusion of dynamic ocean and sea-ice treatments would enable a more complete, and consistent, exploration of the possible climates harbored by terrestrial planets. In such studies, if and where a continent is located will be of particular importance, as continents have been demonstrated to affect oceanic flows \citep{2017arXiv170902051D}. 

Furthermore, a more complete analysis of the acceleration mechanism responsible for maintaining the large-scale flow in the atmosphere, and its response to changes in land surface configuration is required, but well beyond the scope of this work.

Finally, the prediction of potential observable signatures of key transitions between climate states, is a natural follow-up to this work. In particular, if such signatures are predicted and observed, further study may reveal whether it will be possible to infer the presence of a land surface on an exoplanet by observing its effects on the climate. An in-depth description of the prospects for remotely estimating land fraction and location on extrasolar planets is presented in \citet{2012ApJ...756..178A}; here we choose to summarise the main opportunities. A number of studies have now been conducted to investigate the possibility of estimating land fraction and location on exoplanets directly from observations \citep{2009ApJ...700..915C,2010ApJ...715..866F,2010ApJ...720.1333K}. It has been suggested this can be achieved using measurements of reflected visible light, obtained via disc-integrated, time-resolved broadband photometry \citep{2009ApJ...700..915C}, and via a planet's thermal emission spectra \citep{2012ApJ...752....7H}. We comment a little on the former method. This has already been attempted, with relative success, for Earth using data obtained as part of the EPOXI mission \citep{2009ApJ...700..915C}. However, \citet{2010ApJ...723.1168Z} suggest that this technique may not be feasible for the fainter signals received from exoplanets. Another problem is cloud coverage over land, which obscures observations of the surface below. Indeed, the \citet{2010ApJ...720.1333K} surface reconstruction method requires cloud-less skies. This is at odds with permanent cloud coverage near the sub-stellar point on a tidally-locked planet,  meaning that sub-stellar land may be difficult to characterize. In particular, our smaller continents, \emph{B1} and \emph{B2}, retained thick cloud coverage over the entire land surface. The characterization of larger continents provides a slightly more promising target, as the desert climate of these continents promotes reduced cloud coverage.

\section{Conclusions} \label{sec:conclusions}

The key conclusions of this study are:

\begin{enumerate}
	\item The introduction of a sub-stellar land mass serves to reduce the availability of moisture at the sub-stellar point, which decreases both the water vapor greenhouse effect and the cloud radiative effect. This can lead to reduced global mean surface temperatures, chiefly through cooling on the night-side. Day-side surface temperatures exhibit minimal variation because changes in the absorption and reflection of radiation by the day-side atmosphere are roughly reflected by changes in the export of heat from the day-side to the night-side.
	\item Reduced atmospheric water vapor content reduces heat redistribution to the night-side by reducing night-side emissivity. This reduces the night-side top-of-atmosphere infra-red flux and night-side surface temperatures, and exaggerates day-night contrasts in both of these quantities. 
	\item The introduction of land offset to the east of the sub-stellar point can induce a regime change in the large-scale circulation by altering the response of atmospheric waves to the heating perturbation at the sub-stellar point.  In our east-offset continent simulations, two counterrotating mid latitude jets are introduced and the superrotating jet at the equator is weakened. 
	\item Specific to Proxima Centauri B; should the planet reside in a tidally-locked orbit, our results extend previous conclusions regarding its likely habitability to a scenario where both oceans and  a land mass located at its sub-stellar point are present.
\end{enumerate}

\acknowledgments

\emph{Acknowledgements}. We thank Geoffrey K. Vallis for engaging in discussion on  atmospheric waves, which greatly benefited this manuscript. N.T.L. and F.H.L are grateful to the London Mathematical Society for financial support by means of an undergraduate research bursary. I.A.B. and J.M. acknowledge the support of a Met Office Academic Partnership secondment. N.J.M.'s contributions were supported by a Leverhulme Trust Research Project Grant. We acknowledge the use of the MONSooN system, a collaborative facility supplied under the Joint Weather and Climate Research Programme, a strategic partnership between the Met Office and the Natural Environment Research Council. This study contains material produced using Met Office Software. We are grateful to an anonymous referee whose comments were of great value when revising this work.


\end{document}